\RequirePackage{fix-cm}

\documentclass[twocolumn]{svjour3}          
\usepackage[a4paper, total={7.42in, 8in}]{geometry}

\smartqed  

\usepackage{graphicx}
\usepackage{multicol,lipsum}
\usepackage{amsmath}
\usepackage{dcolumn} 
\usepackage{enumitem}
\usepackage{graphicx}
\usepackage{dcolumn}
\usepackage{array,multirow}
\usepackage{bm}
\usepackage{amsmath}
\usepackage{epstopdf}
\usepackage{amsfonts}
\usepackage{amssymb}
\usepackage{mathrsfs}
\usepackage{epsfig}
\usepackage{tabularx}
\usepackage[dvipsnames]{xcolor}
\RequirePackage[colorlinks,citecolor=blue,urlcolor=blue,linkcolor=blue]{hyperref}
\usepackage{cite}
\newcommand{\be}{\begin{equation}}
	\newcommand{\ee}{\end{equation}}
\newcommand{\beq}{\begin{eqnarray}}
	\newcommand{\eeq}{\end{eqnarray}}
	\providecommand{\ed}{\mathrm{d}}

\begin{document}

\title{Classical tests on a charged Weyl black hole: bending of light, Shapiro delay and Sagnac effect}

\author{Mohsen Fathi \and Marco Olivares \and
        J.R. Villanueva 
}

\institute{Mohsen Fathi \at  Instituto de F\'isica y Astronom\'ia, Universidad de Valpara\'iso, Avenida Gran Breta\~na 1111, Valpara\'iso, Chile\\ \email{mohsen.fathi@postgrado.uv.cl}
	\and Marco Olivares \at
              Facultad de
              Ingenier\'{i}a y Ciencias, Universidad Diego Portales, Avenida Ej\'{e}rcito
              Libertador 441, Casilla 298-V, Santiago, Chile\\
              \email{marco.olivaresr@mail.udp.cl}           
           \and
           J.R. Villanueva \at
               Instituto de F\'isica y Astronom\'ia, Universidad de Valpara\'iso, Avenida Gran Breta\~na 1111, Valpara\'iso, Chile\\
               \email{jose.villanueva@uv.cl}
}

\date{Received: date / Accepted: date}

\maketitle

\begin{abstract}
In this paper, we apply the classical test of general relativity on a charged Weyl black hole, the exterior geometry of which is defined by altering the spherically symmetric solutions of the Weyl conformal theory of gravity. The tests are basically founded on scrutinizing the angular geodesics of light rays propagating in the gravitating system caused by the black hole. In this investigation, we bring detailed discussions about the bending of light, together with two other relativistic effects, known as the Shapiro and the Sagnac effects. We show that the results are in good conformity with the general relativistic effects, in addition to giving long-distance corrections caused by the cosmological nature of the background geometry under study.

\end{abstract}

\section{Introduction}

Ever since the late 1990's, the dark matter/dark energy scenario has undergone vigorous efforts to be decoded. The observation of the flat galactic rotation curves \cite{Rubin1980}, the unexpected  gravitational lensing \cite{Massey2010}, and the anti-lensing \cite{Bolejko2013} effects are all related to impacts of an unknown source of mass around the galaxies, the so-called dark matter halo. This is much more complicated when a highly functioning energy source, dark energy, is assumed to be causing the universe's global geometry to expand rapidly. \cite{Riess:1998,Perlmutter:1999,Astier:2012}. These scenarios taken together constitute the most mysterious problems of contemporary cosmology and astrophysics. On the other hand, some believe that these scenarios stem from our lack of knowledge about the behavior of the gravitational field, as a glue to attach each segment of the universe. This opinion has led to a huge number of proposals for extended theories of gravity, mostly including alternatives to Einstein's general relativity. These vary from the most natural ones, i.e., the $f(R)$ theories, to more complicated ones like scalar-tensor, vector-tensor and (non)metric-theories. 

{In recent decades, these proposed gravitational theories have been applied to cosmological models (see Ref.~\cite{Clifton2012pr} for a review), avoiding the need to include dark matter and dark energy. 
In the late 1980's, providing a spherically symmetric vacuum solution to the field equations of the fourth order Weyl conformal gravity (WCG), which had been introduced in 1918 by H. Weyl \cite{Weyl1918mz} and had been revived by R.J. Riegert in 1984 \cite{Riegert1984}, P.D. Mannheim and D. Kazanas showed that the controversial problem of flat galactic rotation curves could be explained by relating it to a specific term included in their solution \cite{Mannheim:1989}. Their solution could also regenerate the usual Schwarzschild-de Sitter spacetime.
This theory is a natural extension of general relativity and proposed as an alternative to the dark matter/dark energy scenario \cite{Mannheim:2005}. Since then, it has been studied from several points of view \cite{Knox:1993fj,Edery:1997hu,Klemm:1998kf,Edery:2001at,Pireaux:2004id,Pireaux:2004xb,Diaferio:2008gh,Sultana:2010zz,Diaferio:2011kc,Mannheim:2011is,Tanhayi:2011dh,Said:2012xt,Lu:2012xu,Villanueva:2013Weyl,Mohseni:2016ylo,Horne:2016ajh,Lim:2016lqv}. This theory was also considered a possibility for understanding the quantum cosmology related to the fluctuations of the early universe \cite{Varieschi2010,Hooft2010a,Hooft2010b,Hooft2011,Varieschi2012isrn,Varieschi2014gerg,Vega2014,Varieschi2014galaxies,Hooft2015}. }

Although it may or may not be the proper alternative theory to general relativity, the Weyl theory of gravity exhibits interesting properties. Most importantly, because of its conformal invariance, it has more conformity with the quantum association of the gravitational field, namely, the graviton.

In this paper, this theory is taken into account while a particular choice for an analytic solution of the extra dark matter-related term in the solution is considered. This choice, obtained in Ref.~\cite{Payandeh:2012mj}, is based on confronting the Mannheim and Kazanas solution with that of the exterior geometry of a charged static spherically symmetric source. In the current investigation, the aforementioned charged source in Weyl gravity constitutes a charged Weyl back hole. We aim at inspecting the behavior of mass-less particles (light beams) that travel on this source. Historically, the inspection of light rays in the gravitational field generated by massive objects formed the foundations of the classical tests of general relativity. Three of these tests are chosen in this paper to be performed on a charged Weyl black hole. The applied method is based on the inspection of the effective potential, inferred from the metric parameters of the exterior geometry and the relevant types of motion are plotted and discussed. This procedure mostly highlights the bending of light and the resultant gravitational lensing around the black hole. Moreover, regarding the relativistic nature of the discussion, the so-called Shapiro delay and the Sagnac effect for light rays in this geometric background are also discussed. 

The work is organized as follows: In Sec.~\ref{ngse}, we examine the calculation of the effective potential and the equations of motion for mass-less particles. {In Sec. \ref{nrg} we calculate the resulting proper and coordinate time for radially in-falling null geodesics}. This is followed by the discrimination of different types of angular motions, developed in terms of their impact parameter relevant to the possible turning points for deflecting trajectories in Sec.~\ref{am}. The fundamental deflection discussed here is then related to the bending of light in  Sec.~\ref{BendingOfLight} and the lens equation is derived. Although light deflection is a rather fundamental concept in relativistic effects, the importance of such derivation is the extension of the deflection distance to the cosmological horizon by the appearance of one extra term in addition to the usual general relativistic term. This focus becomes even more intense by discussing the Shapiro delay in Sec.~\ref{sec:shapiro} for a light ray passing a charged Weyl black hole. {The results demonstrate the long distance effects which are peculiar to the Weyl black hole}. Long distance effects do appear as well for counter-propagating beams in a constant radial distance to the black hole. To highlight this, as the third test, in Sec. \ref{sec:sagnac} we continue with calculating the gravito-magnetic vector potential which generates a phase shift between the beams emitted and absorbed inside a confined rotating apparatus. This shift has a discernible relevancy with the Sagnac and the optical Aharonov-Bohm effects. The results show that the Sagnac time difference is larger for propagation around a Weyl black hole than for a Reissner-Nordstr\"{o}m black hole. We summarize in Sec.~\ref{summ}.

\section{The background}\label{ngse}
{\subsection{Weyl gravity}\label{subsec:WeylGravity}
The Weyl theory of gravity is a theory of fourth order in the metric and is given by the action
\begin{equation}\label{eq:IWeyl}
    I_W=-\mathcal{K}\int{\ed^4x\sqrt{-g}\,\,C_{\mu\nu\rho\lambda}C^{\mu\nu\rho\lambda}},
 \end{equation}
where $g=\mathrm{det}(g_{\mu\nu})$,
\begin{multline}\label{eq:C}
C_{\mu\nu\lambda\rho} = R_{\mu\nu\lambda\rho}-\frac{1}{2}\left(g_{\mu\lambda}
R_{\nu\rho}-g_{\mu\rho}R_{\nu\lambda}-g_{\nu\lambda}R_{\mu\rho}+g_{\nu\rho}R_{\mu\lambda}\right)\\
+\frac{R}{6}\left(g_{\mu\lambda}g_{\nu\rho}-g_{\mu\rho}g_{\nu\lambda}\right)
\end{multline}
is the Weyl conformal tensor and $\mathcal K$ is a coupling  constant. The conformal invariance of the Weyl tensor causes $I_W$ to remain unchanged under the conformal transformation $g_{\mu\nu}(x) = e^{2 \alpha(x)} g_{\mu\nu}(x)$, in which the exponential coefficient indicates local spacetime stretching. The action in Eq.~\eqref{eq:IWeyl} can be rewritten as
\begin{equation}\label{eq:IWeyl-2}
    I_W=-\mathcal{K}\int
\ed^4x
\sqrt{-g}\,\left(R^{\mu\nu\rho\lambda}R_{\mu\nu\rho\lambda}-2R^{\mu\nu}R_{\mu\nu}+\frac{1}{3}R^2\right).
\end{equation}
Since the Gauss-Bonnet term $\sqrt{-g}\,(R^{\mu\nu\rho\lambda}R_{\mu\nu\rho\lambda}-4R^{\mu\nu}R_{\mu\nu}+R^2)$
is a total divergence, it does not contribute
to the equation of motion. We can therefore simplify the action as \cite{Mannheim:1989,Kazanas:1991}
\begin{equation}\label{eq:IWeyl-3}
    I_{W}=-2\mathcal{K}\int{\textmd{d}^4x}\sqrt{-g}\,\,\left(R^{\alpha\beta}R_{\alpha\beta}-\frac{1}{3}R^2\right).
\end{equation}
Applying the principle of least action in the form $\frac{\delta{I_W}}{\delta{g_{\alpha\beta}}} = 0$, leads to the Bach equation $W_{\alpha\beta} = 0$, in which the Bach tensor is defined as
\begin{eqnarray}\label{eq:Bach}
W_{\alpha\beta}&=&\nabla^\sigma\nabla_\alpha R_{\beta\sigma}+\nabla^\sigma\nabla_\beta
R_{\alpha\sigma}-\Box R_{\alpha\beta}-g_{\alpha\beta}\nabla_\sigma\nabla_\gamma
R^{\sigma\gamma}\nonumber\\
&-&2R_{\sigma\beta}
{R^\sigma}_\alpha+\frac{1}{2}g_{\alpha\beta}R_{\sigma\gamma}R^{\sigma\gamma}-\frac{1}{3}\Big(2\nabla_\alpha\nabla_\beta
R-2g^{\alpha\beta}\Box R\nonumber\\
&-&2RR_{\alpha\beta}+\frac{1}{2}g_{\alpha\beta}R^2\Big).
\end{eqnarray}
The general spherically symmetric solution to the Bach equation has been obtained and discussed in Ref.~\cite{Mannheim:1989}, where Mannheim and Kazanas, in addition to recovering all spherically symmetric solutions to Einstein field equations, including the Schwarzschild and Schwarzschild-(Anti-)de Sitter solutions, proposed the possibility of explaining the flat galactic rotation curves, which is claimed to be a significant feature of the dark matter scenario.}

{
The spherically symmetric vacuum solution to the conformal Weyl gravity which, in the usual Schwarzschild coordinates defined in the range $-\infty < t < \infty$,
$r\geq0$, $0\leq\theta\leq\pi$ and $0\leq\phi\leq 2\pi$, is given by the metric \cite{Mannheim:1989}
\begin{equation}
	{\rm d}s^{2}=-B(r)\, {\rm d}t^{2}+\frac{{\rm d}r^{2}}{B(r)}+r^{2}({\rm d}\theta^{2}+\sin^{2}\theta\,
	{\rm d}\phi^{2}), \label{metr}
\end{equation} 
where the lapse function $B(r)$ is defined as
\begin{equation}\label{eq:originalWeylB(r)}
 B(r) = 1-\frac{\xi (2-3\,\xi\,\gamma)}{r}  - 3\,\xi\,\gamma + \gamma r - \kappa r^2, 
\end{equation}
including $\xi$, $\gamma$ and $\kappa$ as the three-dimensional integration constants. In the absence of $\gamma$, the above solution therefore contains the familiar Schwarzschild-de Sitter solution and at distances much smaller than $1/\gamma$, it recovers general relativity. In the presence of a charged static source, the Reissner--Nordstr\"{o}m solution to the theory has been obtained by means of the non-vacuum equation
\begin{equation}\label{eq:W=T}
   W_{\alpha\beta} = \frac{1}{4\mathcal{K}}~T_{\alpha\beta},
\end{equation}
in which $T_{\alpha\beta}$ is the energy-momentum tensor relevant to the vector potential
\begin{equation}\label{eq:vectorPotential}
    A_\alpha = \left(
    \frac{q}{r},0,0,0
    \right),
\end{equation}
where $q$ is the source's electric charge \cite{Mannheim1991}. In the same paper, authors have also dealt with the Kerr and the Kerr--Newman solutions to this fourth order theory of gravity. These charged solutions have been also discussed in Ref.~\cite{Mannheim1991b}.
}

{\subsection{The black hole solution}\label{subsec:Weylblackhole}
The black hole solution studied in the current paper, is indeed constructed on a weak--field method, provided in Ref.~\cite{Payandeh:2012mj}. There, the authors use the Poisson equation for deviations on the background spacetime. By solving Eq.~\eqref{eq:W=T} for the electromagnetic source characterized by Eq.~\eqref{eq:vectorPotential} on the weak field limit, they obtain a solution of the form \cite{Payandeh:2012mj}
\begin{equation}
	B(r)=1-\frac{r^{2}}{\lambda^{2}}
	-\frac{Q^2}{4 r^2}, \label{lapse}
\end{equation}
in which
\begin{equation}
	\label{par1}\frac{1}{\lambda^2}=\frac{3\,\tilde{m}}{\tilde{r}^{3}} +\frac{2\,c_1}{3},\quad Q=\sqrt{2}\,\tilde{q},
\end{equation}
where $\tilde{m}$ and $\tilde{q}$ are the mass and the charge distributed in the spherical system, respectively, and $\tilde{r}$ is its known radius. If the condition $Q<\lambda$ is satisfied, this spacetime allows for two horizons; the event horizon $r_+$ and the cosmological horizon $r_{++}$, located at
\begin{eqnarray}
	\label{rmas}&&r_+=\frac{\lambda}{\sqrt{2}}\sqrt{1-\sqrt{1-\left(\frac{Q}{\lambda}\right)^2}},\\
		\label{rmm}&&r_{++}=\frac{\lambda}{\sqrt{2}}\sqrt{1+\sqrt{1-\left(\frac{Q}{\lambda}\right)^2}}.
\end{eqnarray}
Obviously, the extremal black hole is obtained when $\lambda=Q$, possessing a unique horizon at $r_{ex}=r_+=r_{++}=\lambda/\sqrt{2}$, whereas the naked singularity appears when $\lambda<Q$.}

{It is worth making some clarifications regarding the relevance of the solution given in Eq.~\eqref{lapse} and the well-known static solutions of general relativity. In the absence of electric charge, when the vacuum case in considered, the known radius $\tilde{r}$ changes to the free radial distance $r$. Then, by substituting $3\tilde{m}\rightarrow 2 M$ and $c_1\rightarrow0$ we re-obtain the Schwarzschild spacetime, whereas the Schwarzschild-(Anti-)de Sitter spacetime is regained by letting ${2c_1}\rightarrow\pm\Lambda$ (with $\Lambda$ as the cosmological constant). The corresponding horizons can be then regenerated by solving $B(r)=0$. The relation to the Reissner-Nordstr\"{o}m-(Anti-)de Sitter spacetime, however, requires the imaginary transformation $Q\rightarrow 2 \,i\, Q_0$, in which $Q_0$ is supposed to be the total charge of a spherical massive source. Based on the above types of transformation, it is apparent that the transition from the charged black hole given in Ref.~\cite{Payandeh:2012mj} to the known spherically symmetric spacetimes offered by general relativity, is not trivial. This stems from the mathematical method applied in the derivation of the charged Weyl black hole solution. Let us now continue our discussion on the in-falling geodesics on this black hole.
} \\

\begin{figure}[h!]
	\begin{center}
		\includegraphics[width=80mm]{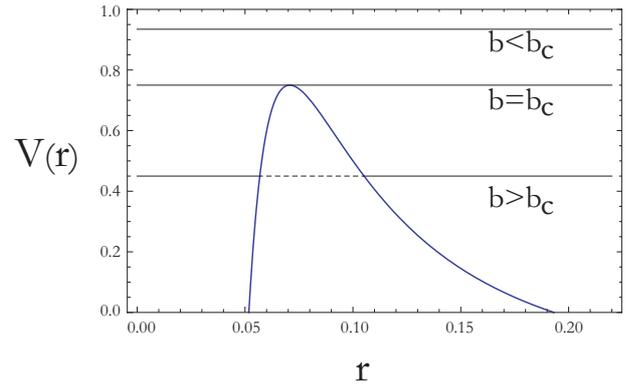}
	\end{center}
	\caption{Plot of the effective potential $V(r)$ versus the radial coordinate $r$, for fixed parameter $L=10^{-1}$, $\lambda=2\times10^{-1}$ and $Q=10^{-1}$ (in arbitrary units).}
	\label{figpot}
\end{figure}

The null geodesic structure of this background can be determined using the standard Lagrangian procedure \cite{Chandrasekhar:579245,Cruz:2004ts,Villanueva:2018kem}, where the (null) Lagrangian associated with the metric (\ref{metr}) reads as
\begin{equation}
	2\mathcal{L}=-B(r)\,\dot{t}^{2}+\frac{\dot{r}^{2}}{B(r)}+r^{2}(\dot{\theta}%
	^{2}+\sin^{2}\theta \dot\phi^2)\equiv 0.
	\label{w.9}
\end{equation}
Here, the dot denotes a derivative with respect to the affine parameter $\tau$ along the geodesics. The equations of motion are then given by 
\be \dot{\Pi}_{\bm\ell} - \frac{\partial \mathcal{L}}{\partial \bm\ell} = 0,
\label{w.10} \ee
where $\Pi_{\bm\ell} = \partial \mathcal{L}/\partial \dot{\bm\ell}$
are the conjugate momenta associated with the generalized coordinates $\bm\ell$.
Clearly, this Lagrangian does not depend on the variables ($t, \phi$) as these are cyclic coordinates. Their conserved conjugate momenta in the invariant plane $\theta =\pi/2$ are
\begin{equation}
	\Pi_{\phi}
	= r^{2} \dot{\phi} = L,\quad
	\textrm{and}\quad 
	\Pi_{t} = -B(r)\, \dot{t} = - E, 
	\label{const}
\end{equation}
where $L$ is the angular momentum (in the units of mass), and $E$ is an integration constants which cannot be considered as the energy because the spacetime is not assymptotically flat.
Furthermore, these two constants of motion allow us to define an impact parameter in terms of the relation $b\equiv\frac{L}{E}$.
Thus, the equations of motion are resumed by the following set of differential equations:       
\begin{eqnarray} \label{rtau}
	&&\left(\frac{{\rm d}r}{{\rm d}\tau}\right)^{2}= E^2-V(r),\\
	\label{rt}
	&&\left(\frac{{\rm d}r}{{\rm d} t}\right)^{2}= B^2(r)\left(1-\frac{V(r)}{E^2}\right),\\
	\label{rphi}
	&&\left(\frac{{\rm d}r}{{\rm d}\phi}\right)^{2}= \frac{r^4}{b^2}\left(1-\frac{V(r)}{E^2}\right),
\end{eqnarray}
where $V(r)$ corresponds to the conformal effective potential defined by
\begin{equation}
	V(r)=L^2\frac{B(r)}{r^{2}},
	\label{poteff}
\end{equation}which is depicted in Fig. \ref{figpot}. This essentially shows same features as that of the de Sitter spacetime in the sense of the existence of two horizons, $r_+$ and $r_{++}$, in this spacetime. We will discuss this potential in more detail in Sec. \ref{am}. In the next section, we begin with scrutinizing the in-falling light beams by letting them be completely radial.

\section{Null radial geodesics}\label{nrg}
\begin{figure}[!h]
	\begin{center}
		\includegraphics[width=80mm]{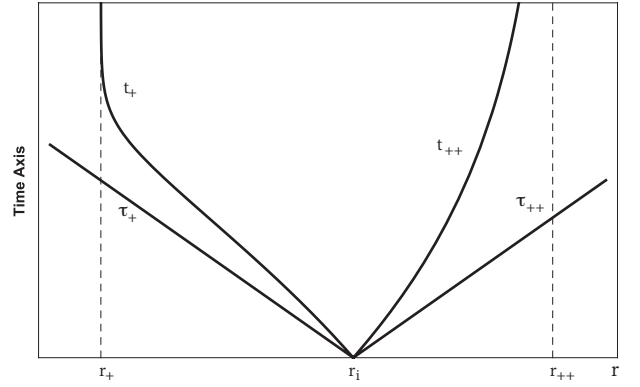}
	\end{center}
	\caption{Temporal behavior for radial null geodesics on the background of a charged Weyl black holes. In the proper system, photons can cross the horizons in a finite time (in accordance with Eq. (\ref{taur})), whereas regarding Eq. (\ref{tr}), an observer at $r_i$ measures an infinite time for $r\rightarrow r_+$ or $r\rightarrow r_{++}$. The same behavior is seen in the study of photon motion in static spherically symmetric spacetimes in the context of general relativity.}
	\label{f2}
\end{figure}

Photons with zero impact parameter perform a radial motion either towards the event horizon or the cosmological horizon. In this case clearly, the effective potential vanishes, such that Eqs. (\ref{rtau}) and (\ref{rt}) become
\begin{equation}
\frac{{\rm d}r}{{\rm d}\tau}=\pm E,\quad \textrm{and}\quad \frac{{\rm d}r}{{\rm d}t}=\pm B(r).
\label{rad1}
\end{equation}
Note that, the sign $+$ ($-$) corresponds to photons falling
onto the cosmological (event) horizon.
Choosing the initial condition $r=r_i$ when $t=\tau=0$ for the photons, a straightforward integration of the first in Eq. (\ref{rad1}) yields
\begin{equation}
\tau(r)=\pm\frac{r-r_i}{E},
\label{taur}
\end{equation}
\noindent which in the proper frame of the photons, indicates that they arrive at the event (cosmological) horizon within a finite proper time. On the other hand, an integration of the second relation in Eq. (\ref{rad1}) leads to
	\begin{equation}
	t(r)=\pm \left[t_{+}(r)+t_{++}(r) \right], \,
	\label{tr}
	\end{equation}
	where
	\begin{equation}
	\label{tm1}
	t_{+}(r)=\frac{\lambda ^2 r_+}{2(r_{++}^2-r_{+}^2)} \ln \left|\frac{r-r_{+}}{r_{i}-r_{+}}
	\,
	\frac{r_i+r_{+}}{r+r_{+}}\right| \,
	\end{equation}
	and
	\begin{equation}
	\label{tm2}
	t_{++}(r)=
	\frac{\lambda ^2 r_{++} }{2(r_{++}^2-r_{+}^2)}\ln  \left|\frac{r_{++}-r_i}{r_{++}-r}
	\,\frac{r_{++}+r}{r_{++}+r_i}\right|\,.
	\end{equation}
Note from Eqs. (\ref{tm1}) and (\ref{tm2}) that the coordinate time in Eq.~(\ref{tr}) diverges for $r\rightarrow r_+$ or $r\rightarrow r_{++}$. Thus, an observer at $r=r_i$ essentially notes the same behavior for photons crossing either of the horizons in a similar manner as in the spherically symmetric spacetimes in the context of general relativity \cite{Chandrasekhar:579245,Cruz:2004ts}. The same holds for uncharged Weyl black holes \cite{Villanueva:2013Weyl} (see Fig.~\ref{f2}). Horizon-crossing, however, can be done in more complex ways once the angular momentum plays its role. This is addressed in the next section.

\section{Angular null geodesics}\label{am}
The angular motion of mass-less particles, whose constants of motion are different from zero is well described by the effective potential (\ref{poteff}). As can be seen in Fig.~\ref{figpot}, the effective potential possesses a maximum at $r_c= Q/\sqrt{2}$, which is independent of $\lambda$. Accordingly, the critical value of the impact parameter is given by 
\begin{equation}\label{eq:bc}
b_c=\frac{\lambda\,Q}{\sqrt{\lambda^2-Q^2}}.
\end{equation}
Comparing the impact parameter of the test particles to this value, we can obtain qualitative descriptions of the angular motions for photons allowed in the exterior spacetime of a charged Weyl black hole. In what follows, we bring detailed discussions about each of these possibilities:

\begin{enumerate}
    	\item \textbf{\emph{Critical Trajectories}}:
	If $b=b_{c}$, an unstable circular orbit of radius $r_{c}$ is allowed as a subset of the null geodesics family.  The proper period in such an orbit is
		\begin{equation}\label{p1}
		T_{\tau}=\pi\,\frac{Q^2}{L},
		\end{equation}
		which is independent of $\lambda$. The coordinate period, however, depends on both $\lambda$ and $Q$ in the form
		\begin{equation}\label{p2}
		T_t=2\pi\,b_c=2\pi\,\frac{\lambda\,Q}{\sqrt{\lambda^2-Q^2}}.
		\end{equation}		
		Thus, photons coming from the initial distance $r_i$, approach asymptotically to the circular orbit at $r_{c}$, according to the appropriate form of Eq. (\ref{rphi}), i.e.,
		\begin{equation}
		\label{criteqmot}\frac{\textrm{d}r}{{\rm d}\phi}=\pm\frac{1}{\sqrt{2}}\frac{(r+r_c)~|r-r_c|}{r_c}.
		\end{equation}
	For those photons coming from outside of $r_c$ ($r_c< r_i<r_{++}$), equation (\ref{criteqmot}) can be recast as
	\begin{equation}
		\label{crfk}\frac{\textrm{d}r}{{\rm d}\phi}=\pm\frac{r^2-r_c^2}{\sqrt{2} \,r_c},
	\end{equation}
	which explains the critical trajectories of the first kind, while the relation
	\begin{equation}
		\label{crsk}\frac{\textrm{d}r}{{\rm d}\phi}=\pm\frac{r_c^2-r^2}{\sqrt{2} \,r_c},
	\end{equation}
	describes those of the second kind for photons initiating their motion from inside of $r_c$ ($r_+ < r_i< r_c$). 	Solutions to these equations can be obtained by direct integration, giving 
		\begin{equation}
	    \label{crt1a}r(\phi)=r_c\,\coth\left(\frac{\phi}{\sqrt{2}} \right),
	\end{equation}
for the first, and
	\begin{equation}
	r(\phi)=r_c\tanh\left(\frac{\phi}{\sqrt{2}}  \right),\label{crt1b}
	\end{equation} 
	for the second kind. In Fig. \ref{fig3}, the critical trajectories \eqref{crt1a} and \eqref{crt1b} have been plotted.
		\begin{figure}[h!]
		\begin{center}
			\includegraphics[width=80mm]{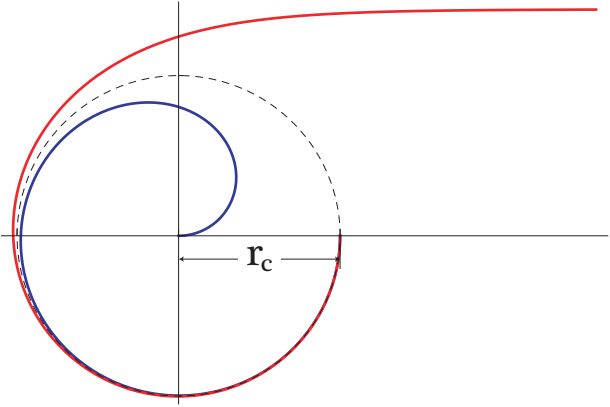}
		\end{center}
		\caption{Critical trajectories of photons with $b=b_c$. Orbits of the first and the second kinds are allowed for test particles that approach by spiraling to the unstable circular orbit at $r=r_c$.}
		\label{fig3}
	\end{figure}

\item \emph{\textbf{Deflection Zone}}. {Light deflection in Weyl gravity, in the context of the Mannheim--Kazanas solution, has been discussed in Refs. \cite{Sultana:2010zz,Sultana2013prd,Sultana2013jcap}. Here, address the same problem, for the charged Weyl black hole under study.} When photons attain the impact parameter $b_{c} <b <\infty$, they are deflected due to the effective potential barrier. Thus, and as in the previous case, they encounter orbits of the first and second kind (OFK and OSK). Photons coming from a finite distance $r_i$ ($r_+ < r_i< r_c$ or $r_c< r_i<r_{++}$)
	to the distance $r=r_f$ or $r=r_{d}$ (which are obtained from the relation $V(r_f)=V(r_d)=E^2$) are then pulled back to either of the two horizons and are in fact deflected. Calculating the turning points, $r_f$ and $r_{d}$, we obtain

	\begin{eqnarray}
	&&r_f=\frac{\beta}{\sqrt{2}}
	\sqrt{1-\sqrt{1-\left(\frac{Q}{\beta}\right)^2} },\label{mr52a}\\
	&&r_d=\frac{\beta}{\sqrt{2}}
	\sqrt{1+\sqrt{1-\left(\frac{Q}{\beta}\right)^2} },\label{mr52b}
	\end{eqnarray}
	where  $\beta$ is the {\it anomalous impact parameter}
		given by
		\begin{equation}\label{mr53}
		\beta=\frac{\lambda\,b}{\sqrt{\lambda^2+b^2 }}.
		\end{equation}
	
	Note that in the limit $b\rightarrow \infty$, the anomalous impact parameter becomes $\beta=\lambda$ and we obtain the identities $r_f(b=\infty)= r_{+}$ and $r_d(b=\infty)= r_{++}$ (see Eqs. (\ref{rmas}), (\ref{mr52a}) and Eqs. (\ref{rmm}), (\ref{mr52b})).	The equations of motion are once again obtained by integrating the general radial relation in Eq. (\ref{rphi}) for both kinds of orbits. To do this, we perform the change of variable $r=\beta \sqrt{u+1/3}$, which generates the equation 
	\begin{equation}
	    \label{angradem}\pm \frac{{\rm d}u}{{\rm d}\phi}=\sqrt{4 u^3-g_2 u-g_3}.
	\end{equation}
This leads to the integrals
\begin{subequations}\label{ofk1}
\begin{align}
     \phi=\int_{u_d}^{u}\frac{{\rm d}u'}{\sqrt{4 u'^3-g_2 u'-g_3}} ~~(\mathrm{with}~ u_d<u),\\
     \phi=\int_{u}^{u_f}\frac{{\rm d}u'}{\sqrt{4 u'^3-g_2 u'-g_3}} ~~(\mathrm{with}~ u_f>u),
\end{align}
\end{subequations}
for OFK and OSK, respectively. The above integrals yield
	\begin{equation}
	    r(\phi)=\beta\sqrt{\frac{1}{3}+\wp(\omega_d-\phi)}\label{mrd}
	\end{equation} for OFK, and
	\begin{equation}
r(\phi)=\beta\sqrt{\frac{1}{3}+\wp(\omega_f+\phi)},\label{mrf}
	\end{equation}	for OSK, in which $\wp(x)\equiv \wp(x; g_2, g_3)$ is the $\wp$-Weierstra\ss\,   function, whose Weierstra\ss\, invariants are given by
		\begin{eqnarray}\label{eq:WeierCoefs}
		&&g_2= \frac{4}{3}-\frac{Q^2}{\beta^2},\label{mr7ca}\\
		&&g_3= \frac{8}{27}-\frac{Q^2}{3\beta^2}.\label{mr7cb}
		\end{eqnarray}
		Furthermore, the phase parameters are given by
		\begin{eqnarray}
		  \label{weiscf}\omega_d&=&\ss\left(\frac{r_d^2}{\beta^2}-\frac{1}{3}\right),\\
		  \label{weiscd}\omega_f&=&\ss\left(\frac{r_f^2}{\beta^2}-\frac{1}{3}\right),
		\end{eqnarray}where $\ss(x)\equiv \ss(x; g_2, g_3)$ is the inverse $\wp$-Weierstra\ss\, function. The qualitative behavior of OFK and OSK is shown in Fig. \ref{fig4}. We should note here that the signature of the above coefficients affects the polynomial on the right hand side of Eq. (\ref{angradem}). Letting $\beta_c = \beta|_{b=b_c}$ we get $\beta_c=Q$ and based on Eqs. (\ref{eq:WeierCoefs}) we have:
		\begin{itemize}
    \item{For $g_2>0$ we have $\bar\beta_2<\beta<\beta_c$,}
    \item{For $g_3>0$ we have $\beta > \bar\beta_3 >\beta_c$,
    in which $\bar\beta_2=\frac{3\beta_c}{2\sqrt{3}}=\sqrt{\frac{2}{3}}\bar\beta_3$.}
\end{itemize}

\begin{figure}[t]
		\begin{center}
			\includegraphics[width=75mm]{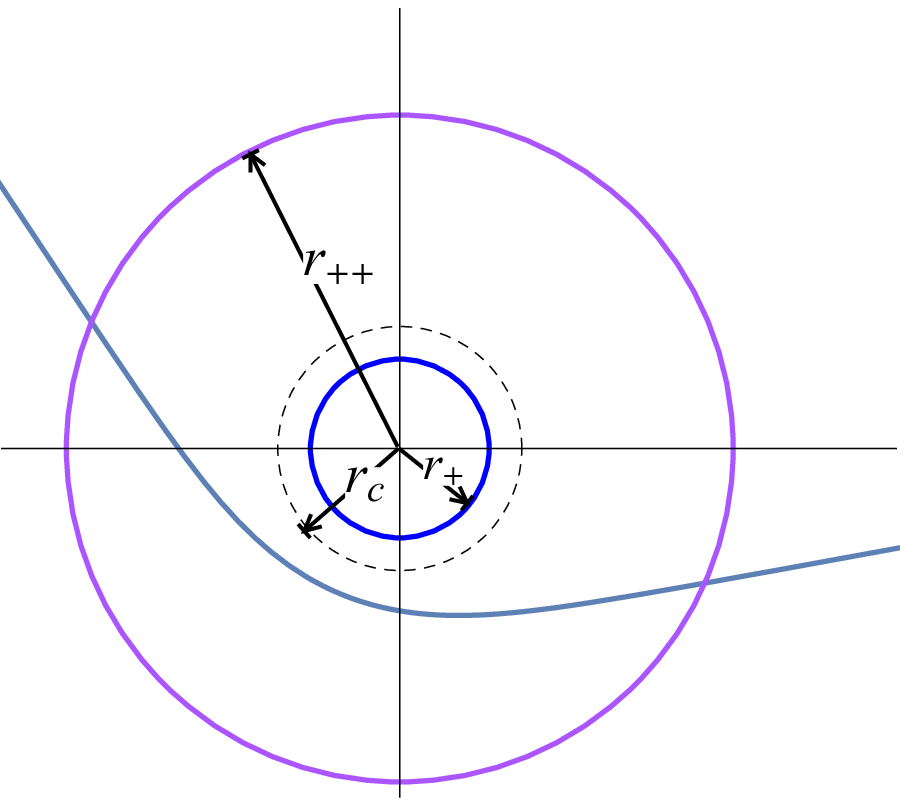}~(a)
			\includegraphics[width=60mm]{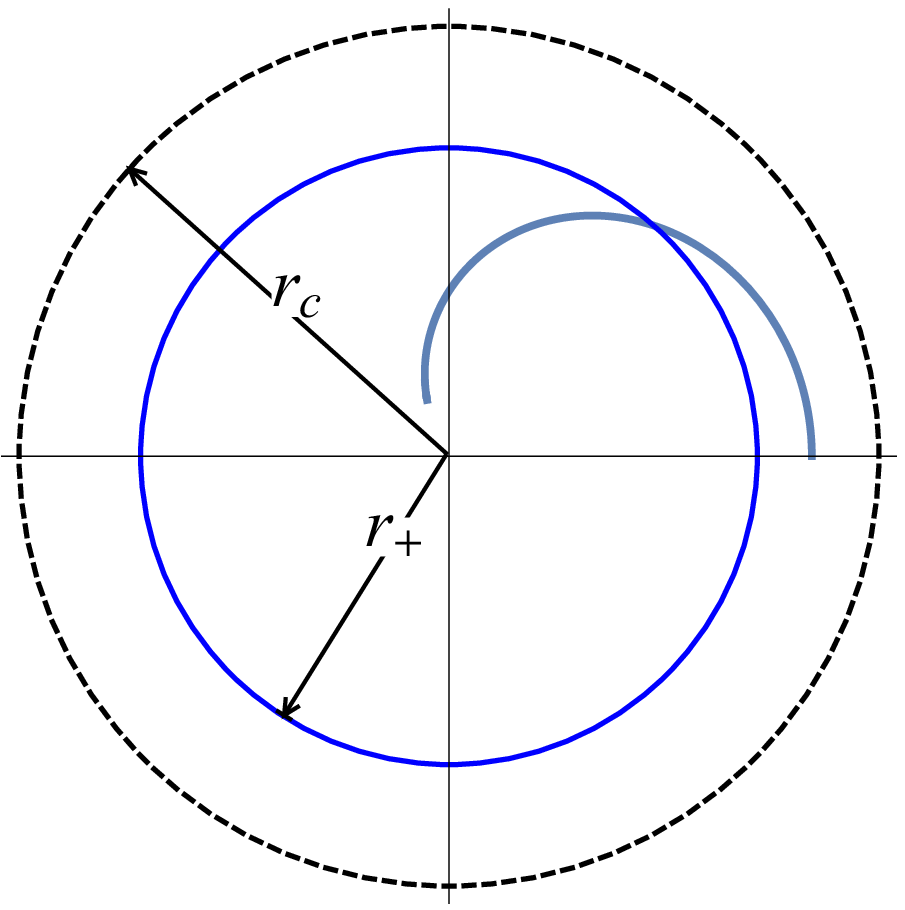}~(b)
		\end{center}
		\caption{ {The deflecting trajectories governed by equations of motion given in Eqs.~(\ref{mrd}) and (\ref{mrf}). The plots demonstrate (a) OFK and (b) OSK. As we can see, the hyperbolic form of OFK allows incoming trajectories to enter the cosmological horizon before their escape to infinity. On the other hand, those that follow OSK, will rapidly enter the event horizon and fall onto the singularity. }}
		\label{fig4}
	\end{figure}
	
{Since we are interested in the region inside the effective potential, we disregard the first case above. We can therefore categorize the following conditions on the coefficients:\\

Condition 1) for $\beta>\bar\beta_3$ we have $g_2>0$ and $g_3>0$.
Condition 2) for $\beta_c<\beta<\bar\beta_3$ we have $g_2>0$ and $|g_3|>0$.\\

It is worth mentioning that, as appears in the decreasing segment of Fig.~\ref{figpot}, the effective potential can change its type of curvature in an inflection point. This appears at the point $r_0$, for which $V''(r_0) = 0$, giving $r_0 = \pm{\sqrt{\frac{5}{6}}}Q$, where
\begin{equation}\label{eq:V0}
    V_0 \equiv V(r_0) = L^2\left(
    \frac{21}{25Q^2}-\frac{1}{\lambda^2}
    \right).
\end{equation}
Moreover, applying the definition in Eq. (\ref{mr53}) to the turning points (where $V(r)=E^2$), we get
\begin{equation}\label{eq:newbeta}
    \frac{1}{\beta^2} = \frac{1}{r_t^2}\left(
    1-\frac{2\bar\beta_3^2}{9r_t^2}
    \right),
\end{equation}
where $r_t$ indicates the turning points, implying that the above relation is valid only on the curve given by the effective potential. From Eq. (\ref{eq:newbeta}) we infer that
\begin{equation}\label{eq:be0vsbeta3}
    \beta_0 = \frac{10\sqrt{2}}{3\sqrt{21}}\bar\beta_3,
\end{equation}
in which $\beta_0 \equiv \beta|_{r=r_0}$. This provides $\beta_0 \approx 1.03 ~\bar\beta_3$. Therefore, the effective potential's value corresponding to $\bar\beta_3$ is larger than $V_0$. However, regarding the small difference between $\bar\beta_3$ and $\beta_0$, geodesics following the OFK described by Eq. (\ref{mrf}) are more likely to fall to bound orbits, as the potential changes from being concave to being convex at $r_0$.
}\\		

	\item \emph{\textbf{Capture Zone}}:
	{Particles with the impact parameter $0<b=b_a<b_c$ will experience an inevitable in-fall onto black hole horizons.} Obviously, the above depends on the initial conditions, specifically on the direction of the velocity at the moment of starting the description of the trajectory.   In both cases, the cross-section is given by \cite{wald}
		\begin{equation}\label{mr51}
		\sigma=\pi\,b_c^2=\frac{\pi\lambda^2 Q^2}{\lambda^2-Q^2}.
		\end{equation}

  {In a similar way as discussed before, we integrate Eq. (\ref{rphi}) to obtain the equation of motion, which reads
	\begin{equation}
	r(\phi)=\beta\sqrt{\frac{1}{3}+\wp(\omega_a+\phi)},
	\label{mr.6c}
	\end{equation}
	where  $\omega_a=\ss(\frac{r_a^2}{\beta^2}-\frac{1}{3})$ is the phase parameter corresponding to the point of approach $r_a$.}
{Note that, depending on the impact parameter, capturing can happen in different ways. As we can see in Fig.~\ref{fig:capture}, for $b<b_c$, the trajectories coming from infinity are captured directly on the event horizon. This is while those with $b=b_c$ follow a spiral-formed trajectory toward the horizon}. \\

Now that the angular motions have been discussed, we can make use of them to relate the features of a charged Weyl black hole to the classical test of general relativity. We start from gravitational lensing.
\begin{figure}[t]
\center{\includegraphics[width=8cm]{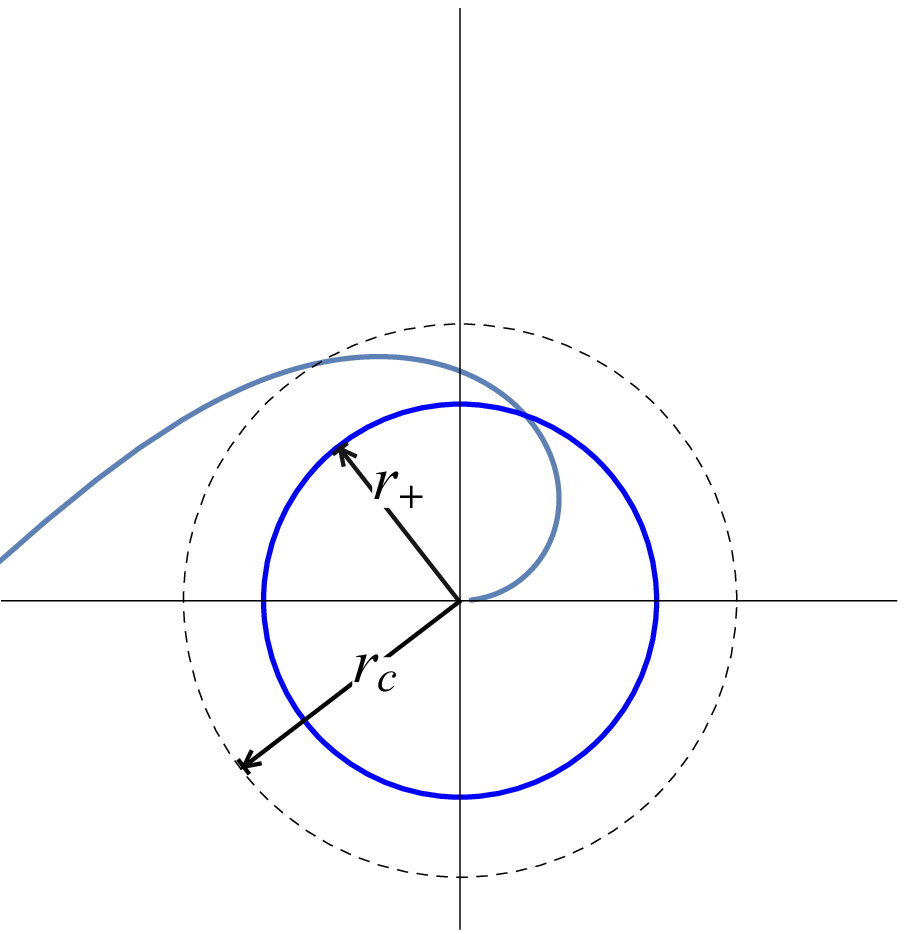}~(a)
\includegraphics[width=6cm]{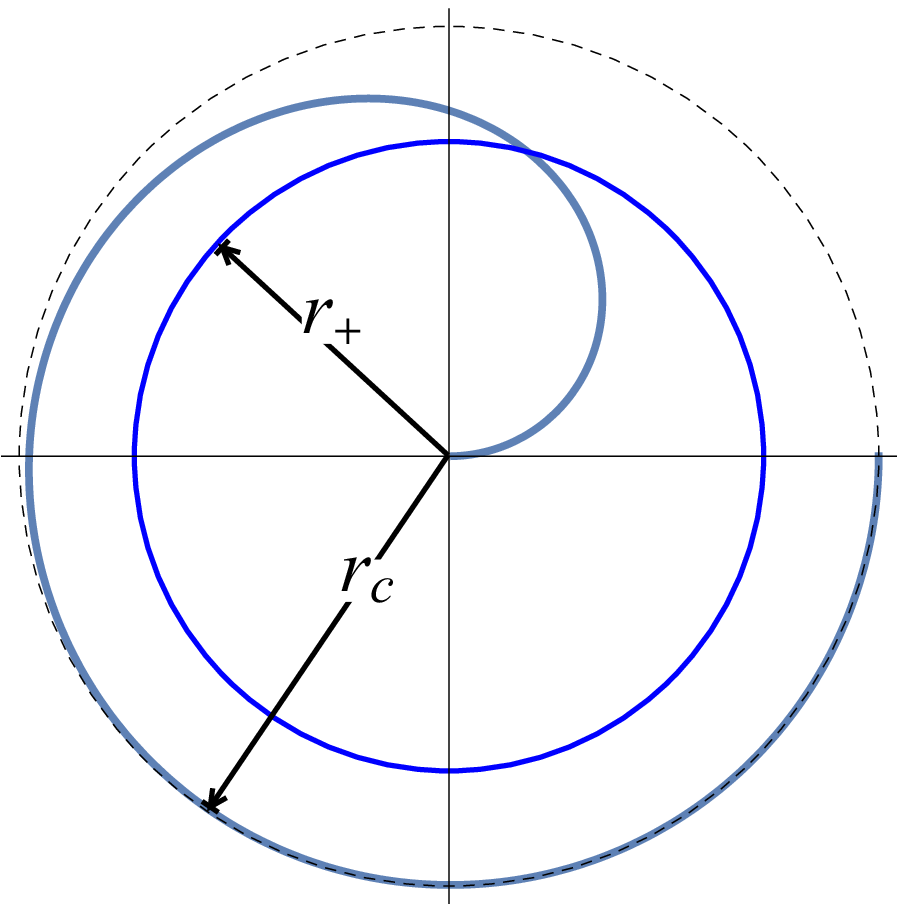}~(b)
\caption{\label{fig:capture} {The capturing process for particles possessing $b\leq b_c$. The figures indicate approaching particles with (a) $b<b_c$ and (b) $b=b_c$. } }}
\end{figure}
\end{enumerate}

\section{Bending of light {and the lens equation}}\label{BendingOfLight}
\begin{figure}
    \centering
    \includegraphics[width=8cm]{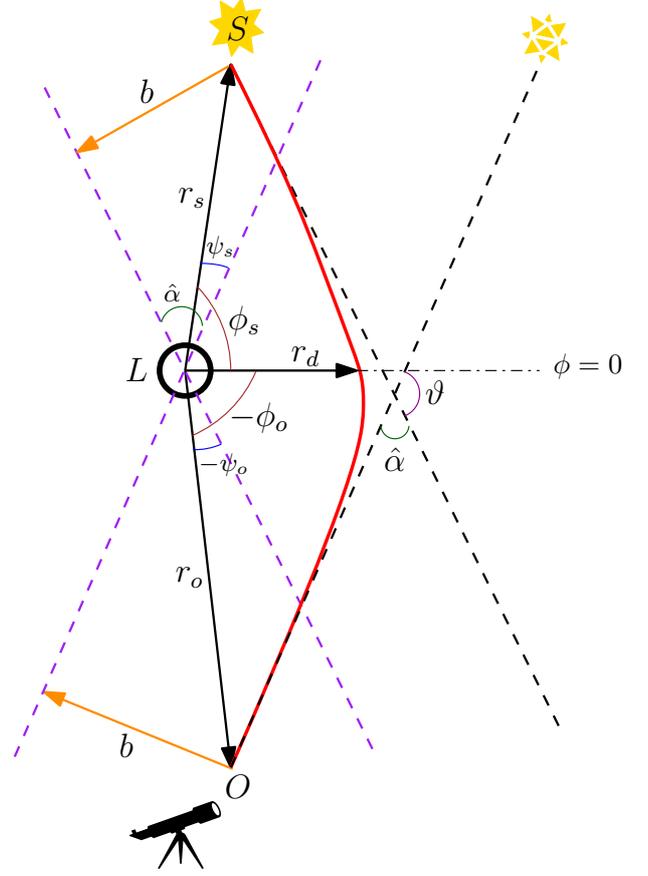}
    \caption{{A schematic illustration of the lensing phenomena. The shortest distance $r_d$ to the lens $L$, has been taken to be the turning point in the OFK, lying on the $\phi=0$ line. The source and the observer are located at $S(r_s, \phi_s,b)$ and $O(r_o,\phi_o,b)$.}} 
    \label{fig:lensing}
\end{figure}

{Regarding the deflection of light in the OFK, gravitational lenses can form. {Gravitational lensing of spherically symmetric black holes has been widely studied in the literature. See the famous paper in Ref.~\cite{Virbhadra2000} for the Schwarzschild black holes and Ref.~\cite{Bozza2010} for more general cases where theoretical aspects of this phenomenon have been developed to compare the predicted higher order images with those of realistic observations. In particular, for charged black holes in the context of Reissner-Nordstr\"{o}m geometry, this effect has been applied in Ref.~\cite{Zhao2016} to study the intrinsic characteristics of the background spacetime. The reader is encouraged to see Ref.~\cite{Treu2015} and references therein to obtain greater insight into the various types of lensing and their applications in astrophysics and cosmology.}  

{Now let us construct the geometry of the problem and apply it to the spacetime under study.} Consider the diagram in Fig.~\ref{fig:lensing}. The source and the observer, characterized by their position, angle and the impact parameter of the light passing them,  are respectively located at $S(r_s, \phi_s, b)$ and $O(r_o, -\phi_o, b)$. The shortest distance $r_d$ to the lens is the turning point given in Eq. (\ref{mr52b}) and $r=r_d$ indicates $\phi=0$. Regarding the figure, we can infer that:
\begin{equation}\label{eq:lens1}
    \vartheta = \phi_s-\psi_s+|\phi_o|-|\psi_o|,
\end{equation}
and $\vartheta + \hat\alpha = \pi$ relates the deflection angle $\hat\alpha$ to the position angles $\phi_o$ and $\phi_s$. It is straightforward to calculate:
\begin{eqnarray}\label{eq:lensing2}
    \psi_s = \hat\alpha - \arcsin(\frac{b}{rs}),\\
    |\psi_o| = \hat\alpha-\arcsin(\frac{b}{r_o}).
\end{eqnarray}
Once again, applying appropriately Eqs. (\ref{rphi}) and (\ref{angradem}), we obtain the angles $\phi_s$ and $\phi_o$ and therefore the lens equation is obtained as:
\begin{eqnarray}\label{eq:lens3}
    \hat\alpha = 
    \arcsin\left(\frac{b}{r_o\,r_s}\left[
    \sqrt{r_o^2-b^2}+\sqrt{r_s^2-b^2}
    \right]\right)
    +2\omega_d \nonumber\\
    - \left[
    \ss\left(\frac{r_s^2}{\beta^2}-\frac{1}{3}\right)
    +
   \left|\ss\left(\frac{r_o^2}{\beta^2}-\frac{1}{3}\right)\right|\right]
    -\pi,
\end{eqnarray}
where $\omega_d$ is the same as that in Eq. (\ref{weiscf}). The above relation, gives the lens equation for light rays passing a charged Weyl black hole.} During the lensing process, as light deflects from the black hole, it experiences a temporal dilation. This causes another important effect which is discussed as the second test in the next section.

\section{Gravitational time delay}\label{sec:shapiro}

\begin{figure}[h!]
	\begin{center}
		\includegraphics[width=70mm]{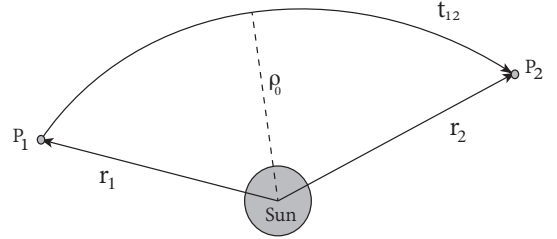}
	\end{center}
	\caption{ {Scheme for the gravitational time delay effect. A light signal is emitted from $P_1$ at $r_1$ to $P_2$ at $r_2$ and returns to $P_1$. Here, $\rho_0$ is the closet approach to the Sun, and $t_{12}$ is the time interval between emission and return of the pulse as measured by a clock on Earth.}}
	\label{ftd}
\end{figure}

{One interesting relativistic effect associated with the propagation of photons, is the apparent delay in the time of propagation for a light signal passing the Sun's proximity. Known as the Shapiro time delay \cite{shapiro64}, this effect is a relevant correction for astronomical observations.} The time delay of radar echoes corresponds to the determination of the time delay of radar signals which are transmitted from the Earth through a region near the Sun to another planet or to a spacecraft, and are then reflected back to Earth (see Fig.~\ref{ftd}). The time interval between emission and return of a pulse as measured by a clock on Earth is given by
\begin{equation}\label{eq:t12def}
t_{12}=2\, [t(r_1,\rho_0)+\, t(r_2,\rho_0)],
\end{equation}
where $\rho_0$ corresponds to the closest proximity to the Sun. 
Returning to Eq. (\ref{rt}):
\begin{equation}\label{eq:rdot}
\dot{r}=\dot{t}\,\frac{{\rm d}r}{{\rm d}t}=\frac{E}{B(r)}\frac{{\rm d}r}{{\rm d}t}=\sqrt{E^2-\frac{L^2}{r^2}B(r)}~.
\end{equation}
Taking into account the fact that at $\rho_0$ the radial velocity $dr/dt$ vanishes, the following relation is obtained:
\begin{equation}\label{TD1}
{\frac{L^2}{E^2} \equiv{b^2}=\frac{\rho_0^2}{B(\rho_0)}}.
\end{equation}
Now, using Eq. (\ref{TD1}) in Eq. (\ref{eq:rdot}), the coordinate time which the light requires to go from $\rho_0$ to $r$ is
\begin{equation}
t(r,\rho_0)=\int_{\rho_0}^r \frac{dr}{B(r)\sqrt{1-\frac{\rho_0^2}{B(\rho_0)}\frac{B(r)}{r^2}}}.
\end{equation}
So, in the first order of corrections, we get
\begin{eqnarray}\label{eq:t(r,rho0)}
t(r, \rho_0)\approx 
\sqrt{r^2-\rho_0^2}+t_Q+t_{\lambda},
\end{eqnarray}
where
\begin{subequations}
\begin{align}
t_Q&=\frac{3\,Q^2}{2\,\rho_0}\sec^{-1}\left({r\over\rho_0}\right)\label{eq:tQ},\\
t_{\lambda}&=\frac{1}{3\lambda^2}
\sqrt{r^2-\rho_0^2}\left[r^2+{\rho_0^2 \over 2}\right]\label{eq:tlambda}.
\end{align}
\end{subequations}
{In the non-relativistic context, light travels in Euclidean space and we can calculate the time interval between emission and reception of the pulse as
\begin{equation}\label{eq:tE12}
    t^E_{12}=2\left(\sqrt{r_1^2-\rho_0^2}+\sqrt{r_2^2-\rho_0^2}\right).
\end{equation}
Therefore, the expected relativistic time dilation in the journey $1\longrightarrow2\longrightarrow1$ can be defined as:
\begin{equation}\label{eq:Deltat}
    \Delta t:=t_{12}-t^E_{12}~,
\end{equation}
which, by exploiting Eqs. (\ref{eq:t12def}) and (\ref{eq:t(r,rho0)}) to (\ref{eq:tE12}), yields
\begin{equation}\label{eq:Deltat2}
    \Delta t = \Delta t_Q + \Delta t_\lambda~,
\end{equation}
where
\begin{subequations}
\begin{align}
\Delta t_Q&=\frac{3\,Q^2}{\rho_0} \left[\sec^{-1}\left({r_1\over\rho_0}\right)+
\sec^{-1}\left({r_2\over\rho_0}\right)\right],\label{eq:DeltatQ1}\\
\Delta t_{\lambda}&=\frac{2}{3\lambda^2}\left[ 
\sqrt{r_1^2-\rho_0^2}\left( r_1^2+{\rho_0^2 \over 2}\right)\right.\nonumber \\
&\left.~~+\sqrt{r_2^2-\rho_0^2}\left( r_2^2+{\rho_0^2 \over 2}\right) \right].\label{eq:Deltatlambda1}
\end{align}
\end{subequations}}
For a round trip in the solar system, we have ($\rho_0<<r_1,r_2$)
\begin{eqnarray}
\Delta t_\odot &\approx& \frac{3\,Q^2}{\rho_0}\left[\sec^{-1}\left({r_1\over\rho_0}\right)+
\sec^{-1}\left({r_2\over\rho_0}\right)\right]\nonumber\\
&&~+\frac{2}{3\lambda^2}\left(r_1^3+r_2^3\right).
\end{eqnarray}
{The above time dilation depends separately on terms relevant to the electric charge and the cosmological constant. However, the closest approach ($\rho_0$) only contributes to the charge-relevant terms, confirming that the electric charge has only short-distance effects, whereas the cosmological term is effective in long distance.} 

The time delay in propagating beams is a completely relativistic effect. In the next section and as the third test, we discuss another specific experiment, relevant to this effect.

\section{The Sagnac effect}\label{sec:sagnac}
The Sagnac effect \cite{sagnac13} is one of the most fascinating classical tests to prove the geometrical nature of gravitation. The study of this phenomena is favored because it can be treated as a formal analogy of the Aharonov-Bohm effect \cite{Sakurai80,Rizzi:2003uc,Rizzi:2003bf,Ruggiero:2005nd,Ruggiero2014epjp}, in the sense that the standard dynamics which raise the natural splitting developed by Cattaneo \cite{Cattaneo1958A,Cattaneo1958,Cattaneo1959,Cattaneo1959b,Cattaneo1959c,Cattaneo:1963aa}, is described in terms of analogue gravito-electromagnetic potentials. Thus, the dynamics of test particles (massive or mass-less), relative to a given time-like congruence $\Gamma$ of the rotating frame of {an ideal} interferometer, can be written in terms of gravito-electromagnetic fields. Therefore, in a rotating frame fixed to the rotating interferometer, the contravariant and covariant components of the unit tangent vector {$\vec\gamma$} to the time-like congruence $\Gamma$ are given by
\begin{eqnarray}
\nonumber \gamma^t&=&1/\sqrt{-g_{tt}},\quad \gamma^i=0,\\\label{congr}
&&\,\\ \nonumber
\gamma_t&=&-\sqrt{-g_{tt}},\quad \gamma_i=g_{it}\gamma^t,
\end{eqnarray}
{where the index $i$ indicates the spatial coordinates.} Here $g_{\mu \nu}$ corresponds to the metric components of the (pseudo--)Riemannian manifold $\mathcal{M}$ in the rotating frame. In this way, the gravito-electromagnetic potentials are defined by \cite{Rizzi2004}
\begin{equation}
    \label{gempot1}\Phi^{G}=-c^2\gamma^t,
\end{equation}and
\begin{equation}
    \label{gempot2}A^{G}_i=c^2\,\frac{\gamma_i}{\gamma_t},
\end{equation}which make it possible to calculate the gravito-magnetic Aharonov-Bohm time difference between the counter-propagating matter or light beams detected by a comoving observer, by means of the relation 
\begin{equation}
    \label{tdse}\Delta \tau =\frac{2\gamma_t}{c^3}\int_{\mathcal{C}}\Vec{A}^G\cdot{\rm d}\Vec{\ell}=\frac{2\gamma_t}{c^3}\int_{\mathcal{S}}\Vec{B}^G\cdot{\rm d}\Vec{a}.
\end{equation}In what follows, we calculate the Sagnac effect using the above expression for the exterior spacetime of a charged Weyl black hole, considering counter-propagating beams on an equatorial plane ($\theta=\pi/2$) along fixed circular trajectories ($r=R$).

\begin{figure}[h!]
		\begin{center}
			\includegraphics[width=80mm]{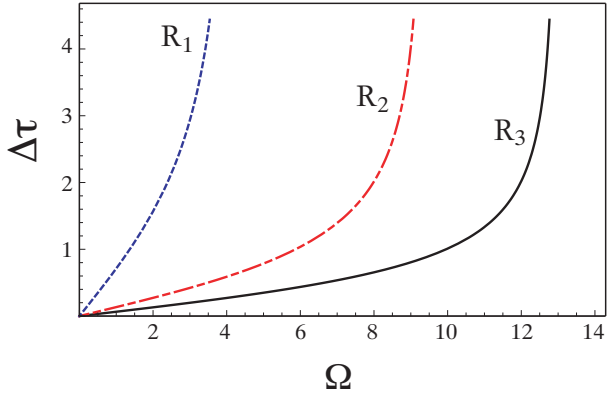}
		\end{center}
		\caption{Time difference $\Delta \tau$ between the counter-propagating beams detected by a comoving observer as a function of the angular velocity $\Omega$, for various separation distances between the source and {an ideal rotating} interferometer. The plots are for $R_1 =7\times 10^7$, $R_2 =3\times 10^7$ and $R_3 =2\times 10^7$ considering $\lambda=2\times 10^{10}$ and $Q=2\times10^7$ (all values are in arbitrary length units).}
		\label{fs1}
	\end{figure}

\begin{figure}[h!]
		\begin{center}
			\includegraphics[width=80mm]{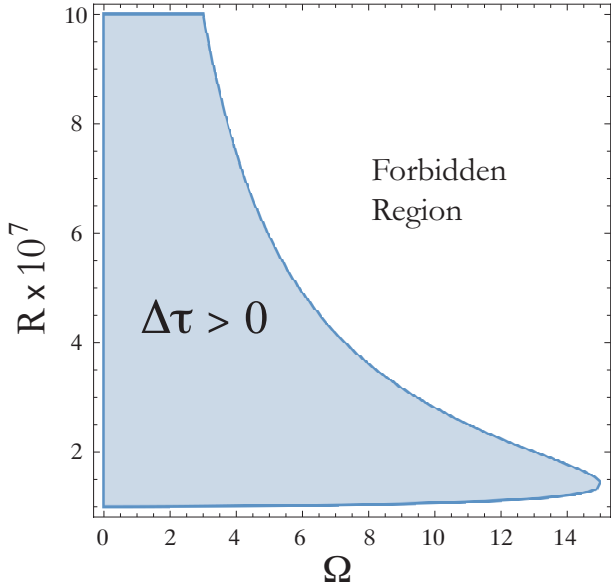}
		\end{center}
		\caption{Region plot for the condition $\Delta \tau>0$ for the separation distances between the source and the interferometer, $R$ and the angular velocity of the comoving observer, $\Omega$, for  $\lambda=2\times 10^{10}$ and $Q=2\times10^7$ (all values are in arbitrary length units).}
		\label{fs2}
	\end{figure}

In order to apply this formalism, let us rewrite the metric (\ref{metr}) by retrieving $c$ in the non-rotating coordinates $x^{\alpha'}=(ct', r', \theta', \phi')$:
\begin{eqnarray}\nonumber
	{\rm d}s^{2}&=&-\left(1-\frac{r'^{2}}{\lambda^{2} }
	-\frac{Q^2}{4 r'^2  }\right)\,c^2 {\rm d}t'^{2}+\frac{{\rm d}r'^{2}}{1-\frac{r'^{2}}{\lambda^{2} }
	-\frac{Q^2}{4 r'^2 }}+\\
	&+&r'^{2}({\rm d}\theta'^{2}+\sin^{2}\theta'\,
	{\rm d}\phi'^{2}). \label{metrpr}
\end{eqnarray}
{The transformation to the local frame of the rotating interferometer (described in $x^\alpha=(c t, r, \theta, \phi)$)  is written as $x^\alpha = {e^\alpha}_{\alpha'} x^{\alpha'}$, in which 
\begin{equation}\label{eq:frameTransformation-e}
    {e^\alpha}_{\alpha'} \equiv \frac{\partial x^\alpha}{\partial x^{\alpha'}} = 
    \left(
\begin{matrix}
   1  & 0 & 0 & 0 \\
   0  & 1 & 0 & 0  \\
   0  & 0 & 1 & 0  \\
   -\Omega & 0 & 0 & 1
\end{matrix}
\right)
\end{equation}
is the frame transformation Jacobian, and $\Omega$ represents the constant angular velocity of the physical system. Thus, we get
\begin{equation}\label{transinterf}
    ct=ct',\,\,r=r',\,\,\theta=\theta',\,\,\phi=\phi'-\Omega t'.
\end{equation}
Applying this, and letting $r=R$ and $\theta=\pi/2$, the line element (\ref{metrpr}) can be recast in $x^\alpha$ as
\begin{eqnarray}\label{eq:rotatingMetric}
 	{\rm d}s^{2}&=&-\left(1-\frac{R^{2}}{\lambda^{2}}
	-\frac{Q^2}{4 R^2}-\frac{R^{2}\Omega^2}{c^{2}}\right)\,c^2 {\rm d}t^{2}+\\
	&-&2\Omega R^{2}{\rm d}\phi {\rm d}t+R^{2}{\rm d}\phi^{2}.
\end{eqnarray} }
Therefore, the components of the vector field ${\vec\gamma(\bm{x})}$, in the rotating frame, are given by
\begin{equation}
    \label{gamJ}\gamma^t=\gamma_J,\quad \gamma_t=-\gamma_J^{-1},\quad \gamma_{\phi}=\frac{\Omega}{\Omega_R}\,R\,\gamma_J,
\end{equation}with
\begin{equation}
    \label{gamJ}\gamma_J=\frac{\Omega_R}{\sqrt{\Omega_0^2-\Omega^2}},
\end{equation}
where $\Omega_0$ is given by
\begin{equation}
    \label{omeg0}\Omega_0\equiv  \sqrt{\Omega^2_R-\Omega_{\lambda}^2-\Omega_{Q}^2},
\end{equation}and
\begin{equation}
    \label{defomeg} \Omega_{R}\equiv \frac{c}{R},\quad \Omega_{\lambda}\equiv \frac{c}{\lambda},\quad \Omega_{Q}\equiv \frac{c \,Q}{2R^2}.
\end{equation}
So, using the above results in Eq. (\ref{gempot2}), we obtain that the only non-zero component of the gravito-magnetic potential is $A_{\phi}^G=-c \Omega R^2 \gamma_J^2$, and the proper time delay between the counter-propagating beams relative to a comoving
observer on the rotating frame is given by
\begin{equation}
    \label{delt1}\Delta\tau=\frac{4\pi}{\Omega_R}\,\frac{\Omega}{\sqrt{\Omega_0^2-\Omega^2}}.
\end{equation}
The variations of this time difference in terms of $\Omega$ have been compared for three different constant spatial separations between the source and the interferometer in Fig.~{\ref{fs1}}. As we can see, the most intense increase in $\Delta\tau$ can happen for smaller $\Omega$ for larger separations. Hence, the same time difference values can be measured in slower rotating interferometers at larger distances from the black hole, as in those with faster rotation at shorter distances. Note that, since $\Delta\tau$ must be positive, an interferometer at a specific distance from the black hole can possess only a definite range of $\Omega$ to work properly. This kind of confinement for the same range of separations and angular velocities used in Fig.~\ref{fs1} has been demonstrated in Fig.~\ref{fs2}.

Essentially, the functional relationship between $\Delta \tau$ and $\Omega$ is the same as that found by Hu {\textit{et al.}} in Ref. \cite{Hu:2006qy}. However, there is a natural shift in the value of the constant $\Omega_0$ compared to the Reissner-Nordstr\"om (RN) case. Clearly, this difference comes from the positivity of the term associated with the electric charge (given by substituting $\Omega_{RN}\rightarrow i \Omega_0$), and also the specific relations to $R$, as the radius of the circular orbits of counter-propagating beams (see Eq. (19) in Ref. \cite{Hu:2006qy}). One important implication of Eq.~(\ref{delt1}) is that, putting aside the Schwarzschild and the electric-charge-associated terms which are common between the RN and the Weyl black holes, the cosmological contribution included in $\Omega_\lambda^2$ results in larger values of $\Delta\tau$ compared to the RN case. This indicates that, unlike the RN case, the Weyl black hole can provide means of measuring the Sagnac effect at large distances.

\section{Summary and outlook}\label{summ}

The geodesic behavior of mass-less particles (light rays) were studied in the exterior geometry of a charged Weyl black hole. The corresponding metric potential contained both an electric charge and cosmological-associated terms. The latter comes into effect particularly in the description of the classical tests done on the black hole. The method was based on analyzing the effective potential in the geodesic equations, in terms of variations in the impact parameter. The impact parameter plays a crucial role in  the determination of the form of in-falling geodesics. This concept was used to distinguish different types of critical, deflected and captured trajectories for photons approaching the black hole's horizons. We showed that the angular equations of motion can be given a well-defined solution based on the {$\wp$-Weierstra\ss}  functions, which are defined in terms of specific invariants. This helped us to obtain reasonable forms of path equations for deflecting trajectories. The first kind of deflecting trajectories in particular, was used further to develop the lensing process in the background geometry under consideration and the lens equation was obtained. Deflecting trajectories were used further to calculate the delay in time for the echo of a light beam passing the black hole. This effect, known as the Shapiro delay, was shown to have long distance effects for the Weyl black hole, according to the cosmological term included in the metric potential. As the final test, we calculated the gravito-magnetic Aharonov-Bohm time difference between the counter-propagating light beams on the black hole's geometry. The results indicated that despite the similarity to the Reissner-Nordstr\"{o}m case, the charged Weyl black hole can make this time difference appear for beams propagating at long distances from the black hole. These long distance effects stem in the extra terms from the cosmological nature of the background spacetime. Based on the results obtained in this investigation, we can infer that a charged Weyl black hole can emulate the features of classical general relativistic black holes, given that the former can present more profound corrections to the classical tests of general relativity, in an apparent relevance with the dark energy related terms in the theory. For this reason, it seems worthy to delve more deeply into the properties of such black holes, because in addition to the reasonable agreement with the classical expectations, they may help us to find out more about the dark energy/dark matter effects on light propagation in strongly gravitating systems. 

\begin{acknowledgements}
	This work was funded by the Comisi\'on Nacional de Investigaci\'on 
Cient\'ifica y Tecnol\'ogica (CONICYT) through DOCTORADO Grant No. 2019-21190382 (MF).
\end{acknowledgements}

\bibliographystyle{spphys} 
\bibliography{Biblio_v1.bib}

\begin{thebibliography}{10}
\providecommand{\url}[1]{{#1}}
\providecommand{\urlprefix}{URL }
\expandafter\ifx\csname urlstyle\endcsname\relax
  \providecommand{\doi}[1]{DOI \discretionary{}{}{}#1}\else
  \providecommand{\doi}{DOI \discretionary{}{}{}\begingroup
  \urlstyle{rm}\Url}\fi

\bibitem{Rubin1980}
V.C. Rubin, W.K. Ford, Jr., N.~Thonnard, Astrophys. J \textbf{238}, 471 (1980).
\newblock \doi{10.1086/158003}

\bibitem{Massey2010}
R.~Massey, T.~Kitching, J.~Richard, Rept. Prog. Phys. \textbf{73}, 086901
  (2010).
\newblock \doi{10.1088/0034-4885/73/8/086901}

\bibitem{Bolejko2013}
K.~Bolejko, C.~Clarkson, R.~Maartens, D.~Bacon, N.~Meures, E.~Beynon, Phys.
  Rev. Lett. \textbf{110}, 021302 (2013).
\newblock \doi{10.1103/PhysRevLett.110.021302}.
\newblock
  \urlprefix\url{https://link.aps.org/doi/10.1103/PhysRevLett.110.021302}

\bibitem{Riess:1998}
A.G. Riess, et~al., Astron. J. \textbf{116}, 1009 (1998).
\newblock \doi{10.1086/300499}

\bibitem{Perlmutter:1999}
S.~Perlmutter, et~al., Astrophys. J. \textbf{517}, 565 (1999).
\newblock \doi{10.1086/307221}

\bibitem{Astier:2012}
P.~Astier, arXiv e-prints arXiv:1211.2590 (2012)

\bibitem{Clifton2012pr}
T.~Clifton, P.G. Ferreira, A.~Padilla, C.~Skordis, Physics Reports
  \textbf{513}(1), 1  (2012).
\newblock \doi{https://doi.org/10.1016/j.physrep.2012.01.001}.
\newblock
  \urlprefix\url{http://www.sciencedirect.com/science/article/pii/S0370157312000105}.
\newblock Modified Gravity and Cosmology

\bibitem{Weyl1918mz}
H.~Weyl, Mathematische Zeitschrift \textbf{2}(3), 384 (1918).
\newblock \doi{10.1007/BF01199420}.
\newblock \urlprefix\url{https://doi.org/10.1007/BF01199420}

\bibitem{Riegert1984}
R.J. Riegert, Phys. Rev. Lett. \textbf{53}, 315 (1984).
\newblock \doi{10.1103/PhysRevLett.53.315}.
\newblock \urlprefix\url{https://link.aps.org/doi/10.1103/PhysRevLett.53.315}

\bibitem{Mannheim:1989}
P.D. Mannheim, D.~Kazanas, Astrophysical Journal \textbf{342}, 635 (1989).
\newblock \doi{10.1086/167623}

\bibitem{Mannheim:2005}
P.D. Mannheim, Prog. Part. Nucl. Phys. \textbf{56}, 340 (2006).
\newblock \doi{10.1016/j.ppnp.2005.08.001}

\bibitem{Knox:1993fj}
L.~Knox, A.~Kosowsky,   (1993)

\bibitem{Edery:1997hu}
A.~Edery, M.B. Paranjape, Phys. Rev. \textbf{D58}, 024011 (1998).
\newblock \doi{10.1103/PhysRevD.58.024011}

\bibitem{Klemm:1998kf}
D.~Klemm, Class. Quant. Grav. \textbf{15}, 3195 (1998).
\newblock \doi{10.1088/0264-9381/15/10/020}

\bibitem{Edery:2001at}
A.~Edery, A.A. Methot, M.B. Paranjape, Gen. Rel. Grav. \textbf{33}, 2075
  (2001).
\newblock \doi{10.1023/A:1013011312648}

\bibitem{Pireaux:2004id}
S.~Pireaux, Class. Quant. Grav. \textbf{21}, 1897 (2004).
\newblock \doi{10.1088/0264-9381/21/7/011}

\bibitem{Pireaux:2004xb}
S.~Pireaux, Class. Quant. Grav. \textbf{21}, 4317 (2004).
\newblock \doi{10.1088/0264-9381/21/18/004}

\bibitem{Diaferio:2008gh}
A.~Diaferio, L.~Ostorero, Mon. Not. Roy. Astron. Soc. \textbf{393}, 215 (2009).
\newblock \doi{10.1111/j.1365-2966.2008.14205.x}

\bibitem{Sultana:2010zz}
J.~Sultana, D.~Kazanas, Phys. Rev. \textbf{D81}, 127502 (2010).
\newblock \doi{10.1103/PhysRevD.81.127502}

\bibitem{Diaferio:2011kc}
A.~Diaferio, L.~Ostorero, V.F. Cardone, J. Cosmol. Astropart. Phys.
  \textbf{1110}, 008 (2011).
\newblock \doi{10.1088/1475-7516/2011/10/008}

\bibitem{Mannheim:2011is}
P.D. Mannheim, Phys. Rev. \textbf{D85}, 124008 (2012).
\newblock \doi{10.1103/PhysRevD.85.124008}

\bibitem{Tanhayi:2011dh}
M.R. Tanhayi, M.~Fathi, M.V. Takook, Mod. Phys. Lett. \textbf{A26}, 2403
  (2011).
\newblock \doi{10.1142/S0217732311036759}

\bibitem{Said:2012xt}
J.L. Said, J.~Sultana, K.Z. Adami, Phys. Rev. \textbf{D85}, 104054 (2012).
\newblock \doi{10.1103/PhysRevD.85.104054}

\bibitem{Lu:2012xu}
H.~Lu, Y.~Pang, C.N. Pope, J.F. Vazquez-Poritz, Phys. Rev. \textbf{D86}, 044011
  (2012).
\newblock \doi{10.1103/PhysRevD.86.044011}

\bibitem{Villanueva:2013Weyl}
J.R. Villanueva, M.~Olivares, J. Cosmol. Astropart. Phys. \textbf{1306}, 040
  (2013).
\newblock \doi{10.1088/1475-7516/2013/06/040}

\bibitem{Mohseni:2016ylo}
M.~Mohseni, M.~Fathi, Eur. Phys. J. Plus \textbf{131}, 21 (2016).
\newblock \doi{10.1140/epjp/i2016-16021-y}

\bibitem{Horne:2016ajh}
K.~Horne, Mon. Not. Roy. Astron. Soc. \textbf{458}(4), 4122 (2016).
\newblock \doi{10.1093/mnras/stw506}

\bibitem{Lim:2016lqv}
Y.K. Lim, Q.h. Wang, Phys. Rev. \textbf{D95}(2), 024004 (2017).
\newblock \doi{10.1103/PhysRevD.95.024004}

\bibitem{Varieschi2010}
G.U. Varieschi, General Relativity and Gravitation \textbf{42}(4), 929 (2010).
\newblock \doi{10.1007/s10714-009-0890-y}.
\newblock \urlprefix\url{https://doi.org/10.1007/s10714-009-0890-y}

\bibitem{Hooft2010a}
G.~{'t Hooft}, arXiv e-prints arXiv:1011.0061 (2010)

\bibitem{Hooft2010b}
G.~{'t Hooft}, arXiv e-prints arXiv:1009.0669 (2010)

\bibitem{Hooft2011}
G.~'t Hooft, Foundations of Physics \textbf{41}(12), 1829 (2011).
\newblock \doi{10.1007/s10701-011-9586-8}.
\newblock \urlprefix\url{https://doi.org/10.1007/s10701-011-9586-8}

\bibitem{Varieschi2012isrn}
G.U. Varieschi, Z.~Burstein, ISRN Astron. Astrophys. \textbf{2013}, 482734
  (2013).
\newblock \doi{10.1155/2013/482734}

\bibitem{Varieschi2014gerg}
G.U. Varieschi, General Relativity and Gravitation \textbf{46}(6), 1741 (2014).
\newblock \doi{10.1007/s10714-014-1741-z}.
\newblock \urlprefix\url{https://doi.org/10.1007/s10714-014-1741-z}

\bibitem{Vega2014}
H.J. {de Vega}, N.G. {Sanchez}, arXiv e-prints arXiv:1304.0759 (2013)

\bibitem{Varieschi2014galaxies}
G.U. Varieschi, Galaxies \textbf{2}(4), 577 (2014).
\newblock \doi{10.3390/galaxies2040577}.
\newblock \urlprefix\url{https://www.mdpi.com/2075-4434/2/4/577}

\bibitem{Hooft2015}
G.T. {Hooft}, arXiv e-prints arXiv:1410.6675 (2014)

\bibitem{Payandeh:2012mj}
F.~Payandeh, M.~Fathi, Int. J. Theor. Phys. \textbf{51}, 2227 (2012).
\newblock \doi{10.1007/s10773-012-1102-1}

\bibitem{Kazanas:1991}
D.~Kazanas, P.D. Mannheim, Astrophysical Journal Supplement Series \textbf{76},
  431 (1991).
\newblock \doi{10.1086/191573}

\bibitem{Mannheim1991}
P.D. Mannheim, D.~Kazanas, Phys. Rev. D \textbf{44}, 417 (1991).
\newblock \doi{10.1103/PhysRevD.44.417}.
\newblock \urlprefix\url{https://link.aps.org/doi/10.1103/PhysRevD.44.417}

\bibitem{Mannheim1991b}
P.D. Mannheim, Annals of the New York Academy of Sciences \textbf{631}(1), 194
  (1991).
\newblock \doi{10.1111/j.1749-6632.1991.tb52643.x}.
\newblock
  \urlprefix\url{https://nyaspubs.onlinelibrary.wiley.com/doi/abs/10.1111/j.1749-6632.1991.tb52643.x}

\bibitem{Chandrasekhar:579245}
S.~Chandrasekhar, \emph{The mathematical theory of black holes}.
\newblock Oxford classic texts in the physical sciences (Oxford Univ. Press,
  Oxford, 2002).
\newblock \urlprefix\url{https://cds.cern.ch/record/579245}

\bibitem{Cruz:2004ts}
N.~Cruz, M.~Olivares, J.R. Villanueva, Class. Quant. Grav. \textbf{22}, 1167
  (2005).
\newblock \doi{10.1088/0264-9381/22/6/016}

\bibitem{Villanueva:2018kem}
J.R. Villanueva, F.~Tapia, M.~Molina, M.~Olivares, Eur. Phys. J. \textbf{C78},
  10 (2018).
\newblock \doi{10.1140/epjc/s10052-018-6328-5}

\bibitem{Sultana2013prd}
C.~Cattani, M.~Scalia, E.~Laserra, I.~Bochicchio, K.K. Nandi, Phys. Rev. D
  \textbf{87}, 047503 (2013).
\newblock \doi{10.1103/PhysRevD.87.047503}.
\newblock \urlprefix\url{https://link.aps.org/doi/10.1103/PhysRevD.87.047503}

\bibitem{Sultana2013jcap}
J.~Sultana, Journal of Cosmology and Astroparticle Physics \textbf{2013}(04),
  048 (2013).
\newblock \doi{10.1088/1475-7516/2013/04/048}.
\newblock \urlprefix\url{https://doi.org/10.1088%2F1475-7516%2F2013%2F04%2F048}

\bibitem{wald}
R.M. Wald, \emph{{General relativity}} (Chicago Univ. Press, Chicago, IL,
  1984).
\newblock \urlprefix\url{https://cds.cern.ch/record/106274}

\bibitem{Virbhadra2000}
K.S. Virbhadra, G.F.R. Ellis, Phys. Rev. D \textbf{62}, 084003 (2000).
\newblock \doi{10.1103/PhysRevD.62.084003}.
\newblock \urlprefix\url{https://link.aps.org/doi/10.1103/PhysRevD.62.084003}

\bibitem{Bozza2010}
V.~Bozza, General Relativity and Gravitation \textbf{42}(9), 2269 (2010).
\newblock \doi{10.1007/s10714-010-0988-2}.
\newblock \urlprefix\url{https://doi.org/10.1007/s10714-010-0988-2}

\bibitem{Zhao2016}
F.~Zhao, J.~Tang, F.~He, Phys. Rev. D \textbf{93}, 123017 (2016).
\newblock \doi{10.1103/PhysRevD.93.123017}.
\newblock \urlprefix\url{https://link.aps.org/doi/10.1103/PhysRevD.93.123017}

\bibitem{Treu2015}
T.~Treu, R.S. Ellis, Contemporary Physics \textbf{56}(1), 17 (2015).
\newblock \doi{10.1080/00107514.2015.1006001}.
\newblock
  \urlprefix\url{https://www.tandfonline.com/doi/abs/10.1080/00107514.2015.1006001}

\bibitem{shapiro64}
I.I. Shapiro, Phys. Rev. Lett. \textbf{13}, 789 (1964).
\newblock \doi{10.1103/PhysRevLett.13.789}.
\newblock \urlprefix\url{https://link.aps.org/doi/10.1103/PhysRevLett.13.789}

\bibitem{sagnac13}
G.~Sagnac, C. R. Acad. Sci. \textbf{157}, 708 (1913).
\newblock \urlprefix\url{https://ci.nii.ac.jp/naid/10021107346/en/}

\bibitem{Sakurai80}
J.J. Sakurai, Phys. Rev. D \textbf{21}, 2993 (1980).
\newblock \doi{10.1103/PhysRevD.21.2993}.
\newblock \urlprefix\url{https://link.aps.org/doi/10.1103/PhysRevD.21.2993}

\bibitem{Rizzi:2003uc}
G.~Rizzi, M.L. Ruggiero, Gen. Rel. Grav. \textbf{35}, 1745 (2003).
\newblock \doi{10.1023/A:1026053828421}

\bibitem{Rizzi:2003bf}
G.~Rizzi, M.L. Ruggiero, Gen. Rel. Grav. \textbf{35}, 2129 (2003).
\newblock \doi{10.1023/A:1027345505786}

\bibitem{Ruggiero:2005nd}
M.L. Ruggiero, Gen. Rel. Grav. \textbf{37}, 1845 (2005).
\newblock \doi{10.1007/s10714-005-0190-0}

\bibitem{Ruggiero2014epjp}
M.L. Ruggiero, A.~Tartaglia, The European Physical Journal Plus
  \textbf{129}(6), 126 (2014).
\newblock \doi{10.1140/epjp/i2014-14126-y}.
\newblock \urlprefix\url{https://doi.org/10.1140/epjp/i2014-14126-y}

\bibitem{Cattaneo1958A}
C.~Cattaneo, Il Nuovo Cimento \textbf{10}, 318 (1958)

\bibitem{Cattaneo1958}
C.~Cattaneo, Il Nuovo Cimento \textbf{10}(2), 318 (1958).
\newblock \doi{10.1007/BF02732487}.
\newblock \urlprefix\url{https://doi.org/10.1007/BF02732487}

\bibitem{Cattaneo1959}
C.~Cattaneo, Il Nuovo Cimento \textbf{13}(1), 237 (1959).
\newblock \doi{10.1007/BF02727548}.
\newblock \urlprefix\url{https://doi.org/10.1007/BF02727548}

\bibitem{Cattaneo1959b}
C.~Cattaneo, Il Nuovo Cimento \textbf{11}(5), 733 (1959).
\newblock \doi{10.1007/BF02732334}.
\newblock \urlprefix\url{https://doi.org/10.1007/BF02732334}

\bibitem{Cattaneo1959c}
C.~Cattaneo, Atti Accad. Naz. Lincei, VIII. Ser., Rend., Cl. Sci. Fis. Mat.
  Nat. \textbf{27}, 54 (1959)

\bibitem{Cattaneo:1963aa}
C.~Cattaneo, G.~Caricato, U.~Roma., \emph{Introduzione alla teoria einsteiniana
  della gravitazione} (Eredi Virgilio Veschi, Roma, 1963)

\bibitem{Rizzi2004}
G.~Rizzi, M.L. Ruggiero, \emph{The Relativistic Sagnac Effect: Two Derivations}
  (Springer Netherlands, Dordrecht, 2004), pp. 179--220.
\newblock \doi{10.1007/978-94-017-0528-8_12}.
\newblock \urlprefix\url{https://doi.org/10.1007/978-94-017-0528-8_12}

\bibitem{Hu:2006qy}
P.H. Hu, Y.J. Wang, Chin. Phys. Lett. \textbf{23}, 2341 (2006).
\newblock \doi{10.1088/0256-307X/23/8/103}

\end{thebibliography}

\end{document}